\documentclass[12pt,preprint]{emulateapj}
\usepackage{graphicx,longtable,float}
\usepackage{amssymb,txfonts,color}
\usepackage[dvipdfm,colorlinks=true,linkcolor=blue,urlcolor=blue,citecolor=blue]{hyperref}
\usepackage{rotating,graphicx}
\usepackage{longtable}
\usepackage{ulem}
\usepackage{float}

\newcommand{\mbh}{M_{\rm BH}}
\newcommand{\msun}{{\rm M}_{\sun}}
\newcommand{\ledd}{L_{{\rm Edd}}}

\newcommand{\ergs}{\rm \,erg\,s^{-1}}

\newcommand{\lx}{{L_{\rm X}}}
\newcommand{\lr}{{L_{\rm R}}}
\newcommand{\chandra}{{\textit{Chandra}}}

\newcommand{\xmm}{{\textit{XMM-Newton}}}

\shorttitle{Fundamental Plane of quiescent AGNs}
\shortauthors{Xie \& Yuan}

\begin{document}
\title{Fundamental Plane of Black Hole Activity in Quiescent Regime}

\author{Fu-Guo Xie\altaffilmark{1} and Feng Yuan\altaffilmark{1}}
\altaffiltext{1}{Key Laboratory for Research in Galaxies and Cosmology, Shanghai Astronomical Observatory, Chinese Academy of Sciences, 80 Nandan Road, Shanghai 200030, China; fgxie@shao.ac.cn, fyuan@shao.ac.cn}

\begin{abstract}
A correlation among the radio luminosity ($L_{\rm R}$), X-ray luminosity ($L_{\rm X}$), and black hole mass ($M_{\rm BH}$) in active galactic nuclei (AGNs) and black hole binaries is known to exist and is called the ``Fundamental Plane'' of black hole activity. \citet{YC2005} predicts that the radio/X-ray correlation index, $\xi_{\rm X}$, changes from $\xi_{\rm X}\approx 0.6$ to $\xi_{\rm X}\approx 1.2-1.3$ when $L_{\rm X}/L_{\rm Edd}$ decreases below a critical value $\sim 10^{-6}$. While many works favor such a change, there are also several works claiming the opposite. In this paper, we gather from literature a largest quiescent AGN (defined as $L_{\rm X}/L_{\rm Edd}\la 10^{-6}$) sample to date, consisting of $75$ sources. We find that these quiescent AGNs follow a $\xi_{\rm X}\approx 1.23$ radio/X-ray relationship, in excellent agreement with the Yuan \& Cui prediction. The reason for the discrepancy between the present result and some previous works is that their samples contain not only quiescent sources but also ``normal'' ones (i.e., $L_{\rm X}/L_{\rm Edd}\ga 10^{-6}$). In this case, the quiescent sources will mix up with those normal ones in $L_{\rm R}$ and $L_{\rm X}$. The value of $\xi_{\rm X}$ will then be between $0.6$ and $\sim1.3$, with the exact value being determined by the sample composition, i.e., the fraction of the quiescent and normal sources. Based on this result, we propose that a more physical way to study the Fundamental Plane is to replace $L_{\rm R}$ and $L_{\rm X}$ with $L_{\rm R}/L_{\rm Edd}$ and $L_{\rm X}/L_{\rm Edd}$, respectively.
\end{abstract}


\section{Introduction}
\label{sec:intro}

There are many observational evidences in black hole (BH) X-ray binaries (BHBs) and active galactic nuclei (AGNs) for the coupling between the collimated relativistic jet and the accretion flow. One such evidence comes from \citet{Corbel2003} and \citet{Gallo2003}, who discovered a remarkably tight correlation between radio (monochromatic, $\lr = \nu L_\nu$ at, e.g. 5 or 8.5 GHz) and X-ray (at 2--10 keV; hereafter $\lx$) luminosities in BHBs during their hard states (see \citealt{Corbel2013} for latest summary). This correlation was later extended to include low-luminosity AGNs (LLAGNs). With the impact of black hole mass $\mbh$ taken into account, \citet[][hereafter M03]{Merloni2003} found that, $\log\lr=0.6\log\lx + 0.78\log\mbh+ 7.33$, with a scatter of $\sigma_{\rm R} = 0.88 $ dex. Here (and throughout this paper) the luminosities and BH mass are, respectively, in units of $\ergs$ and $\msun$. We refer this relationship as the {\it original/standard} M03  ``Fundamental Plane'' (hereafter FP) of back hole activity (for later work, see e.g., \citealt{Falcke2004, Kording2006, Merloni2006, Wang2006, Panessa2007, Li2008, Gultekin2009}, hereafter G09; \citealt{Plotkin2012, Younes2012, Dong2015, Panessa2015, Fan2016, Liu2016, Nisbet2016}).

\citet{YCN2005} have proposed a coupled accretion--jet model for LLAGNs and BHBs (see \citealt{Yuan2014} for a review). In this model, unless the system is extremely faint, the thermal gas in a hot accretion flow is responsible for the X-ray emission and the relativistic power-law distribution electrons in a jet produce the radio emission. \citet[][hereafter YC05]{YC2005} show that the FP can be explained naturally in this model (seel also M03; \citealt{Heinz2003}; \citealt{Heinz2004}; \citealt{XY2016}), i.e. it is a direct consequence of a tight relationship between mass inflow/accretion rate and mass ejection rate (YC05; \citealt{XY2016}). The scatter of the correlation, on the other hand, may reflects the (combination of) effects of other parameters, e.g., the intrinsic variability in radio and X-rays, the BH spin \citep{Miller2009, Narayan2012}, the strength of magnetic field \citep{BZ77, Sikora2007}, the Doppler beaming effect (\citealt{Li2008}, but see \citealt{VVF2013}), the angular momentum of the accreting gas \citep{Cao2016} and the environment \citep{VVF2013}.

After the discovery of the original M03 FP, several notable complexities are revealed over the past decade. First, AGNs with different radio loudness seem to follow relationships that are different in both normalization and correlation slope (e.g., \citealt{Wang2006, Li2008, deGasperin2011}).
 Secondly, long-term quasi-simultaneous monitoring on individual sources find that, some sources do not follow the slope predicted by the original M03 FP during their fluctuations in luminosities, and they can be classified as outliers (for BHBs, see e.g. \citealt{Xue2007, Coriat2011}, and \citealt{Corbel2013} for a recent summary; for AGNs, see e.g. \citealt{Bell2011, King2013, Xie2016}). These outliers, {\it individually}, seem to follow a hybrid radio/X-ray correlation, i.e. $\lr\propto L_{\rm X}^{\sim 1.3}$ when $\lx$ is high (see also \citealt{Gallo2012, Dong2014, Panessa2015, Qiao2015}), $\lr\propto L_{\rm X}^{\sim 0}$ when $\lx$ is moderately low, and can recover the original M03 FP when $\lx$ is much lower. The hybrid correlation is most evident in BHB H1743--322 \citep{Coriat2011} and LLAGN NGC 7213 \citep{Bell2011, Xie2016}.

The third complexity, which is the focus of this work, is whether or not the AGNs with extremely low luminosities, i.e. the so-called ``quiescent'' AGNs (defined as sources with $L_{\rm X}/L_{\rm Edd}\la 10^{-6}$, here $\ledd=1.3\times10^{46}\ \mbh/10^8\ \msun\ \ergs$ is the Eddington luminosity), follow the original M03 FP. YC05 shows that the answer should be ``no''. The reason is that in the accretion--jet scenario, the origin of the X-ray emission in quiescent AGNs is different to that of normal LLAGNs (defined as sources with $L_{\rm X}/L_{\rm Edd}\ga 10^{-6}$), i.e. it originates from jet instead of hot accretion flow (YC05). Physically, this is because the both the hot accretion flow and the jet emit X-ray emission, but the dependence of X-ray radiation from the hot accretion flow on the accretion rate is less sensitive compared to that from the jet. Thus with the decrease of accretion rate (or equivalently luminosity), the radiation from the jet will catch up with that from the accretion flow and become dominant below a critical luminosity $L_{\rm X,crit}/L_{\rm Edd}\approx 10^{-6}$ (see also \citealt{Fender2003}). Moreover, because of the change of the X-ray origin from a hot accretion flow to a jet, YC05 predicts that the quiescent accretion systems will follow a steeper relationship between radio and X-rays (see also \citealt{Heinz2004, Gardner2013}), i.e., the FP of faint accretion systems is revised to (YYH09), $\log\lr=1.23\log\lx + 0.25 \log\mbh -13.45$. Hereafter we will refer to this relationship as the YC05 FP.

We note that the value of the critical luminosity $L_{\rm X,crit}$ depends on detailed parameters that control the properties of accretion flow and/or jet. The $L_{\rm X,crit}$ value obtained in YC05 is for the ``general'' sources. If for some reason (e.g., Doppler beaming) the radiation from the jet is very strong in some sources, the critical luminosity can become significantly larger. This is the case of radio loud sources (e.g., \citealt{Wang2006, Li2008, deGasperin2011}). In this work, we aim at ``general'' sources.

The YC05 predictions have been confirmed by many observational and theoretical works (see review by Yuan \& Narayan 2014), both of AGNs (e.g., \citealt{Pellegrini2007, Wu2007, Wrobel2008, Yuan2009}, hereafter YYH09; \citealt{deGasperin2011, Younes2012, Li2016}) and BHBs (e.g., \citealt{Pszota2008, Plotkin2013, Reynolds2014, Xie2014, YL2015}). For example, YYH09 did a statistical analysis based on a sample consisting of 22 quiescent AGNs. They found $\xi_{\rm X}=1.22\pm0.02$, in excellent agreement with YC05 prediction.

However, there are also claims in literature on an universal FP extending down to quiescent systems, without any steepening pattern. For example, the three BHBs with radio and X-ray observations during their quiescent states are claimed to follow the original M03 FP at low X-ray luminosities \citep{Gallo2006, Gallo2014, Corbel2008, Corbel2013}. For AGNs, recently \citet[][hereafter DW15]{Dong2015} selected from the flux-limited Polmar survey a sample of 73 AGNs (Sgr A* also included), and combine with a large sample of data points of BHBs, to investigate the FP {\it jointly} from sub-Eddington to quiescent systems. They claim that those with $\lx \lesssim L_{\rm X,crit}$ (24 AGNs under a detailed definition of $L_{\rm X,crit}$, cf. DW15) seem to be roughly consistent the original M03 FP. As discussed in detail later in Sections\ \ref{sec:otherworks} \& \ref{sec:jointfit}, we argue their analysis and conclusion have some problems.

In this work, the FP of quiescent systems is re-examined. Because the aim is to check the YC05 prediction, following YYH09, we exclude normal LLAGNs and focus on quiescent AGNs only. In order to reduce the contamination of the host galaxy, we restrict ourselves to observations that have sufficiently high spatial resolution and sensitivity. This paper is organised as follows. We describe our sample compilation (75 sources in total, to our knowledge the largest quiescent AGN sample to date) in Section\ \ref{sec:sample}. We then introduce the statistical analysis method in Section\ \ref{sec:method}. Subsequently we present the fitting results in Section\ \ref{sec:results}. Sections\ \ref{sec:discussion} and \ref{sec:summary} are devoted to discussions and a brief summary respectively.

\section{The Sample of quiescent/faint AGNs}
\label{sec:sample}

We gather from literature a sample of quiescent/faint AGNs with measurements in their black hole mass, nuclei radio (at 5 GHz) and X-ray (2-10 keV) luminosities. Since most of our targets are weak nuclei at the center of normal galaxies, contamination of emission from their host galaxy, i.e. the radio emission due to residual star formation or supernovae remnants in the nuclei region of the host galaxy, or extended/elongated radio emission from mini-lobes if the jet is spatially resolved (cf. discussions in \citealt{Nyland2016}), could be of vital importance. In order to discriminate these possible contaminations, we select data which are observed with arcsec (or even higher, e.g., \xmm\ and \chandra\ for X-rays, and Very Large Array [VLA] for radio.) spatial resolution. For sources that have several observations, we prefer data with higher spatial resolution. Moreover, for sources with extended radio morphologies, we only consider the nuclei component, i.e. the emission of the compact jet. Considering the strong fluctuations in AGN activity, we argue that only this compact radio component relates directly to the {\it current} nuclei activity (shown in X-rays) of the AGNs. We also note that for the X-rays of NGC~3115, only the compact component is adopted \citep{Wong2014}.

The BH mass in these systems is mostly derived based on the empirical $\mbh$--$\sigma$ relationship (cf. \citealt{Kormendy2013} for review of various ways to estimate $\mbh$), whose uncertainty is normally $\sigma_{\rm M} \approx 0.3$. More reliable $\mbh$ measurements, such as those derived from kinematics or reverberation-mapping, will be adopted if exist, cf. references in Table \ref{tab1} for details. Besides, in order to convert the observed flux to luminosity, the luminosity distance $d_{\rm L}$ should be given in advance. Because of the luminosity constraint, most of the sources in our sample have $d_{\rm L}\lesssim 50$ Mpc. Consequently, most of our sources have redshift-independent distance measurements. For most of these sources, the distances adopted in this work are from \citet{Tully2013}, where we select distances constrained through the surface brightness fluctuation (SBF) method. For the rest few sources that lack redshift-independent distance measurements, we derive $d_{\rm L}$ from redshift based on the Planck2015 flat cosmology \citep{Planck2016}, with $H_0 = 67.8\ {\rm km\ s^{-1}\ Mpc^{-1}}$ and $\Omega_M=0.308$.

\begin{figure}[htb]
\centering{
\includegraphics[scale=0.45]{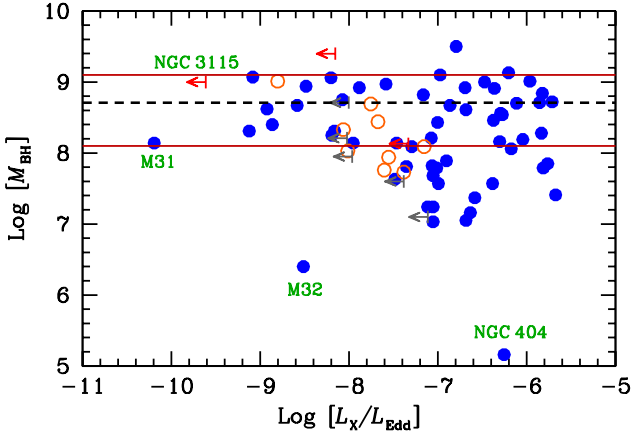}}
\caption{Distribution of black hole mass (in unit of $\msun$) as a function of X-ray Eddington ratio ($\lx/\ledd$). The two red solid curves mark the boundaries of the (quasi-)$\mbh$-free subsample, i.e. $\log \mbh = 8.1, 9.1$, and the black dashed curve shows the average BH mass of this subsample, $ \log \langle\mbh\rangle=8.71$. The blue filled circles represent sources with firm detections in $\lr$ and $\lx$, the orange open circles and red arrows show sources whose $\lr$ and $\lx$ are, respectively, upper limit constraints. The dark grey arrows are sources whose $\lr$ and $\lx$ are both upper limit constraints. Several notable sources are labelled in green.}\label{fig:sample}
\end{figure}

The main selection criteria of our sample compilation comes from the X-ray luminosity constraint, i.e. we require the X-ray Eddington ratio $\lx/\ledd \lesssim 10^{-6}$, a critical luminosity below which the accretion--jet model predicts a jet origin of the X-ray emission (YC05; YYH09). Considering the uncertainties in the measurements of $\mbh$, practically this criteria is slightly weakened to $\lx/\ledd < 10^{-5.7}$ (cf. the X-axis range of Fig.\ \ref{fig:sample}). There is one notable exception -- we exclude from our sample Sgr A*, whose $\lx/\ledd \sim 10^{-11}$ (in its quiescent state, cf. Fig.\ \ref{fig:fpedd}). High-resolution radio observations suggest the non-existence of elongated/collimated jet in this system during its quiescent state (at $\sim 10\ R_{\rm s}$ level, where $R_{\rm s}$ is the Schwardschild radius of BH, see e.g., \citealt{Shen2005, Doe2008}),\footnote{Note that, as an independent approach, the jet in quiescent state of Sgr A* is recently ruled out by the reliable measurement of the Faraday Rotation Measure at submillimeter wavelength \citep{Li2014}.}
and theoretically the low-frequency radio emission (e.g. at 5 GHz) originates from the relativistic power-law distribution electrons within the hot accretion flow itself \citep{Yuan2003}, instead of the jet. Moreover, the X-ray emission in Sgr A* is dominated by diffuse gas around $\sim 10^5\ R_{\rm s}$ (e.g. \citealt{Baganoff2003, Wang2013}), unlike that in other normal LLAGNs where it originates from nuclei $<50\ R_{\rm s}$ regions (e.g., \citealt{Fabian2009, Emm2014}). Detailed discussion on Sgr A*, including its flare state, will be given later in Section\ \ref{sec:sgra}.

As listed in Table \ref{tab1} and shown in Fig.\ \ref{fig:sample}, our final sample includes 75 sources, mainly selected from previous compilations, e.g., M03; \citet{Nagar2005, Hardcastle2009, Ho2009, HGFS2009}; YYH09; \citet{Pellegrini2010}; DW15 and \citet{Nyland2016}. As shown in Fig.\ \ref{fig:sample}, to our knowledge this is the largest quiescent AGN sample to date, in which 5 sources have $\lx/\ledd\lesssim10^{-9}$. In this sample, 58 sources have firm detections in both radio and X-rays (blue filled circles in Fig.\ \ref{fig:sample}). 8 sources have upper limit constrains in radio but firm detections in X-rays (orange open circles in Fig.\ \ref{fig:sample}, and orange arrows in the rest figures), while 4 sources have firm detections in radio but only upper limit constrains in X-rays (red arrows in Fig.\ \ref{fig:sample}). The rest 5 sources only have upper limit constrains in both radio and X-rays (grey arrows in Fig.\ \ref{fig:sample}). As shown in Fig.\ \ref{fig:sample}, this quiescent AGN sample covers a large dynamical range in both radio and X-ray Eddington ratios, i.e. $10^{\sim -14} < \lr/\ledd < 10^{\sim -7}$ (cf. Fig.\ \ref{fig:fpedd} below) and $10^{\sim -10} < \lx/\ledd < 10^{\sim -5.7}$. The BH mass of most sources is between $10^{\sim 7.5}\ \msun$ and $10^{\sim 9}\ \msun$, with a few exceptions, e.g., $\mbh = 10^{5.16}\ \msun$ for NGC~404 \citep{Ho2009} and $\mbh = 10^{6.4}\ \msun$ for M32 \citep{Kormendy2013}. We emphasis that the nuclei of M31, with $\lr/\ledd = 10^{-14.0}$ and $\lx/\ledd = 10^{-10.2}$ \citep{Garcia2010}, represents the faintest (in Eddington unit) source in our sample.

With the fact that the LLAGNs are expected to vary moderately on timescale of months to years \citep{Ho2001, Ho2008}, the uncertainties of the data mainly comes from the non-simultaneity between radio and X-rays (Note that additionally, the time delay between the two wavebands should be corrected, cf. Section\ \ref{sec:questions}). Considering the systematic and the observational uncertainties, we take isotropic uncertainties with $\sigma_{\rm R}=\sigma_{\rm X}=\sigma_{\rm M}=0.3$ dex, following M03 and G09.

\section{Fitting Method}\label{sec:method}

We consider the following linear (in logarithmic space) relationship among three quantities, e.g., $\lr$ (in unit of $\ergs$), $\lx$ (in unit of $\ergs$) and $\mbh$ (in unit of $\msun$),
\begin{equation}
\log \lr = \xi_{\rm X} \log \lx + \xi_{\rm M} \log \mbh + c.\label{eq:fitform}
\end{equation}

For our statistical analysis, we adopt the Markov chain Monte Carlo (MCMC) Bayesian analysis (cf. e.g. \citealt{Kelly2007, FM2013}, hereafter Bayesian approach.) to derive the best-fit parameters and their corresponding uncertainties (see e.g., \citealt{Plotkin2012}). For this approach, we take the \texttt{Python} routine \texttt{Emcee} \citep{FM2013} ver. 2.2.1, which is based on the affine invariant MCMC ensemble sampler method \citep{Goodman2010}. Besides, we assume the intrinsic scatter of the FP is of Gaussian distribution. Moreover, since $\lx$ (or $\lx/\ledd$), $\lr$ (or $\lr/\ledd$) and $\mbh$ are symmetric physical quantities during the modelling, we in practice set the model prior probability function $p(\xi_{\rm X}, \xi_{\rm M}, c)$ to \citep{VanderPlas2014},
\begin{equation}
p(\xi_{\rm X}, \xi_{\rm M}, c) = (1+\xi_{\rm X}^2)^{-3/2}\, (1+\xi_{\rm M}^2)^{-3/2}
\end{equation}

We note that the multivariate correlation coefficients shown in several influential works (among others see e.g., M03 and G09) are derived through minimization of the following statistics (hereafter least $\chi^2$ approach),
\begin{equation}
\chi^2 = \Sigma\ {(\log \lr - \xi_{\rm X} \log \lx - \xi_{\rm M} \log \mbh - c)^2\over \sigma_{\rm R}^2 + \xi_{\rm X}^2 \sigma_{\rm X}^2 +  \xi_{\rm M}^2 \sigma_{\rm M}^2}.
\end{equation}
Although this method can provide reasonable regression coefficients (cf. \citealt{Fasano88}), there are indeed some concerns (e.g., \citealt{Plotkin2012}). For the sake of direct comparison to those previous works, we also provide results under this approach, but our discussions will be mainly based on results derived through the Bayesian regression analysis. Technically we use the \texttt{Python} routine \texttt{kmpfit}\footnote{https://github.com/josephmeiring/kmpfit} of the \texttt{Kapteyn} package ver. 2.3 \citep{Terlouw2012}, in which the \texttt{C} implementation of \texttt{mpfit} \citep{Markwardt2009} is adopted.

\section{Results}
\label{sec:results}

In this section, we present the numerical results based on different approaches. Section\ \ref{sec:allsample} considers the sample consisting of all the quiescent AGNs (72 sources), where both the Bayesian and the least $\chi^2$ methods are adopted (cf. Section\ \ref{sec:method} above). Section\ \ref{sec:mbhfreesample} represents the results based on a sub-sample with similar black hole mass (42 sources). This is because we want to focus on the value of $\xi_{\rm X}$ so we hope to eliminate any possible contamination by the black hole mass. We find a convergency in both methods, i.e. $\xi_{\rm X}\sim 1.2$--1.4.

\begin{figure*}[htb]
\centering{\includegraphics[scale=0.75]{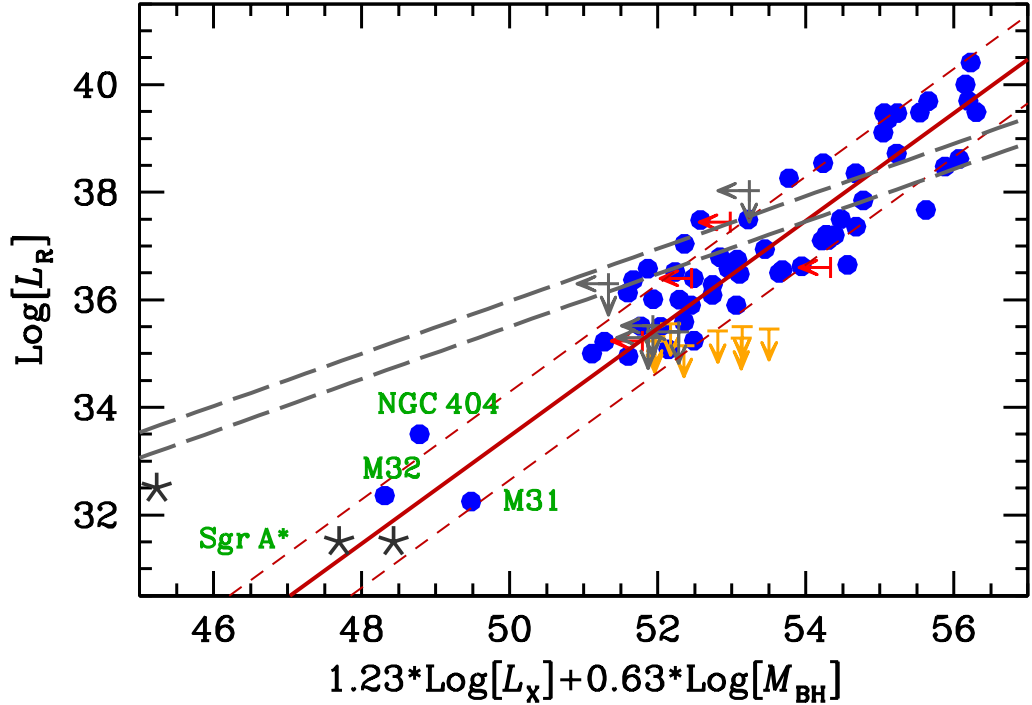}}
\caption{The Fundamental Plane of quiescent AGNs. The red solid and dashed curves represent the Bayesian regression analysis result (cf. Eq.\ \ref{eq:fpfaint}) and its scatter, respectively. For comparison, we show the original M03 FP by the two grey long-dashed curves, with $\mbh$ fixed to $10^9\ \msun$ (upper) and $10^8\ \msun$ (lower). The meaning of the symbols is the same as that in Fig.\ \ref{fig:sample} (note that the orange arrows here are the same as the orange circles in Fig.\ \ref{fig:sample}). Several notable sources are labelled in green. Additionally, Sgr A* in its quiescent and flare states, is shown by black asterisks for the purpose of comparison, see Sec.\ \ref{sec:sgra} for details.}\label{fig:fp}
\end{figure*}

\begin{figure}[htb]
\centering{\includegraphics[scale=0.45]{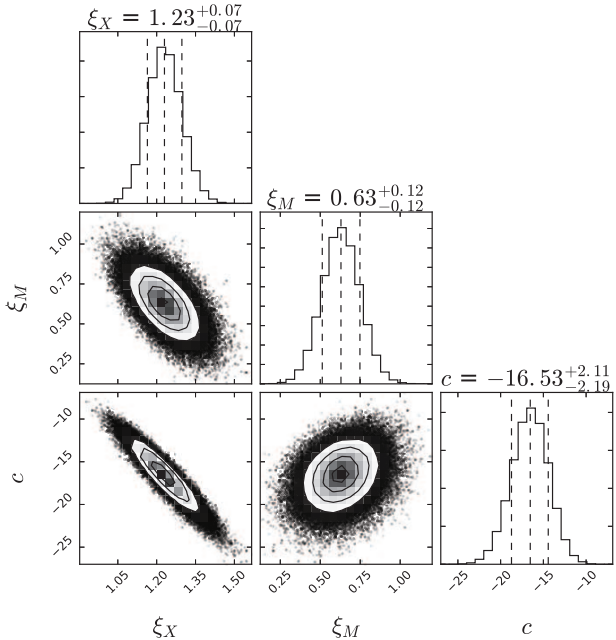}}
\caption{The one and two dimensional projections of the posterior probability distributions of the fitting parameters based on Bayesian MCMC analysis of the whole quiescent AGN sample, cf. Eq.\ \ref{eq:fpfaint}. The three dashed curves mark the location of the best-fit value, and the corresponding 1$\sigma$ uncertainties.}\label{fig:fitpar}
\end{figure}

\subsection{The FP of quiescent AGNs}
\label{sec:allsample}

Here we investigate the linear relationship among $\log (\lr)$, $\log (\lx)$ and $\log(\mbh)$, following the methodologies described in Section\ \ref{sec:method}. We first consider a sub-sample of 58 sources, which have firm detections in both radio and X-rays (hereafter firm-detection subsample). We find that, they follow a FP with parameters constrained as,
\begin{equation}
\xi_{\rm X} = 1.09^{+0.07}_{-0.07}, \hspace{0.5cm} \xi_{\rm M} = 0.70^{+0.12}_{-0.11}, \hspace{0.5cm} c = -11.61^{+2.19}_{-2.25},
\end{equation}
under the Bayesian regression analysis, and
\begin{equation}
\xi_{\rm X} = 1.12^{+0.07}_{-0.07}, \hspace{0.5cm} \xi_{\rm M} = 0.82^{+0.11}_{-0.11}, \hspace{0.5cm} c = -13.60^{+2.29}_{-2.29},
\end{equation}
under the least $\chi^2$ approach. Obviously the firm-detection subsample shows a steeper radio/X-ray correlation slope compared to that of the original FP.

We then analyse the whole quiescent AGN sample. Under the Bayesian approach, the fitting parameters read,
\begin{equation}
\xi_{\rm X} = 1.23^{+0.07}_{-0.07}, \hspace{0.5cm} \xi_{\rm M} = 0.63^{+0.12}_{-0.12}, \hspace{0.5cm} c = -16.53^{+2.11}_{-2.19},  \label{eq:fpfaint}
\end{equation}
with the intrinsic scatter in $\log \lr$ to be $\sigma_{\rm R} = 0.81$ dex,
and under the least $\chi^2$ method the fitting parameters read,
\begin{equation}
\xi_{\rm X} = 1.30^{+0.07}_{-0.07}, \hspace{0.5cm} \xi_{\rm M} = 0.57^{+0.10}_{-0.10}, \hspace{0.5cm} c = -18.53^{+2.19}_{-2.19}, \label{eq:fpfaint_chi2}
\end{equation}
with the intrinsic scatter in $\log \lr$ to be $\sigma_{\rm R} = 0.82$ dex.

The radio/X-ray correlation slope derived above (e.g. Eqs.\ \ref{eq:fpfaint}\&\ref{eq:fpfaint_chi2}) is clearly different, at $>6\sigma$ confidence level, from that of the original M03 FP, whose $\xi_{\rm X}\approx 0.6\pm0.1$ (see also \citealt{Corbel2013} for summary of this relationship in BHBs). This FP is in good agreement (in $\xi_{\rm X}$) with the prediction of YC05 and the result of YYH09 based on a limited quiescent AGN sample. We show the Bayesian regression analysis result (cf. Eq.\ \ref{eq:fpfaint}) as the red solid curve in Fig.\ \ref{fig:fp}. The two red dashed curves represent the intrinsic $1\sigma$ scatter of the best-fit result. For comparison, we also show the original M03 FP by the two black long-dashed curves, where the black hole mass is fixed to $10^8\ \msun$ and $10^9\ \msun$, respectively, for the lower and the upper curves. Fig.\ \ref{fig:fitpar} shows one and two dimensional projections of the posterior probability distributions of Bayesian MCMC fitting parameters, where the three dashed curves mark the location of the best-fit values, and the corresponding 1$\sigma$ uncertainties (cf. Eq.\ \ref{eq:fpfaint}).

From Fig.\ \ref{fig:fp}, there are several points worth further emphasis. First, considering the moderately large scatter in the observational data, the deviation to the original M03 FP (cf. the long-dashed curves) is most evident in M31 (see also Fig.\ \ref{fig:fp_dw} below and \citealt{Wu2013}), which has lowest $\lx/\ledd$ and $\lr/\ledd$ in our sample. The other sources, on the other hand, are in rough agreement with the original M03 FP (see also Fig.\ \ref{fig:fp_dw} below). In this sense, we urge to devote more efforts on sources whose X-ray Eddington ratio $\lx/\ledd \la 10^{-9.5}$. These sources are of crucial importance in confirming the deviation to the original FP, as well as the existence of a new YC05-type FP in these faint/quiescent AGNs. Secondly, the radio/X-ray correlation slope $\xi_{\rm X}$ in quiescent AGNs can be constrained from a sample which includes quiescent AGNs only. If instead the sample also includes numerous brighter sources, the new steep correlation will become invisible. We will discuss these two issues further in Section\ \ref{sec:otherworks}.

\LongTables
\begin{deluxetable}{lcccl}[htb]
\tablewidth{0pt}
\tablecaption{Table 1. Observational data of quiescent AGNs}
\tablehead{
\colhead{Sources}  & \colhead{Log($M_{\rm{BH}}$)}  &
\colhead{Log($L_{\rm R}$)}   & \colhead{Log($L_{\rm x}$)}  &
\colhead{Refences$^a$}\\
\colhead{} & \colhead{($\msun$)} & \colhead{($\ergs$)}  &
\colhead{($\ergs$)}  & \colhead{}}
\startdata
3C 31       &   $8.70$  &   $39.48$  &   $40.70$  & m: 1; rx: 2\\ 
3C 66B     &   $8.84$  &   $40.00$  &   $41.13$  & m: 1; rx: 2\\
3C 83.1B  &   $9.01$  &   $39.49$  &   $41.16$  & m: 1; rx: 2\\ 
3C 338     &   $8.92$  &   $39.47$  &   $40.34$  & m: 1; r: 3$^b$; x: 2 \\
3C 449     &   $8.54$  &   $39.11$  &   $40.38$  & m: 1; rx: 2\\
3C 465     &   $9.13$  &   $40.44$  &   $41.07$  & m: 1; rx: 2\\
M31   &   $8.14$  &   $32.25$  &   $36.06$  & m: 4; rx: 5\\ 
M32   &   $6.40$  &   $32.36$  &   $36.00$  & m: 6; rx: 7\\ 
M81   &   $7.85$  &   $37.20$  &   $40.20$  & m: 8; rx: 9\\ 
M84   &   $8.97$  &   $38.54$  &   $39.50$  & mx: 10; r: 11, 12\\ 
M87   &   $9.5$  &   $39.70$  &   $40.82$  & m: 13, rx: 12\\ 
NGC 0404   &   $5.16$  &   $33.5$  &   $37.02$  & mrx: 14,1\\
NGC 0474   &   $7.73$  &   $<35.55$  &   $38.46$  & mr: 12; x: 10\\
NGC 0507   &   $8.91$  &   $37.67$  &   $40.66$  & mrx: 14, 1\\
NGC 0524   &   $8.94$  &   $36.75$  &   $38.57$  & mr: 12; x: 10\\
NGC 0821   &   $8.21$  &   $<35.40$  &   $<38.30$  & mrx: 10, 12\\
NGC 1399   &   $8.7$  &   $<38.03$  &   $<38.82$  & mx: 10; r: 15$^b$\\
NGC 2768   &   $8.82$  &   $37.50$  &   $39.77$  & mrx: 12; x: 16\\
NGC 2778   &   $7.16$  &   $35.50$  &   $38.64$  & mrx:12; x: 16\\
NGC 2787   &   $8.14$  &   $36.52$  &   $38.30$  & mx: 14, 1 r: 17\\ 
NGC 2841   &   $8.31$  &   $36.00$  &   $38.26$  & mx: 14, r: 17\\
NGC 3115   &   $9.00 $  &   $35.23$  &   $<37.50$  & mr: 18; x: 19\\
NGC 3226  &   $8.06$  &   $37.21$  &   $40.00$  & mx: 14, r: 17\\
NGC 3245  &   $8.21$  &   $36.94$  &   $39.25$  & mx: 14, r: 17\\
NGC 3377  &   $8.25$  &   $35.08$  &   $38.17$  & mrx: 10\\
NGC 3379  &   $8.62$  &   $36.01$  &   $37.81$  & mrx: 14, 10\\ 
NGC 3384  &   $7.03$  &   $35.22$  &   $38.09$  & mr: 12, 15; x: 10\\ 
NGC 3414  &   $8.67$  &   $36.65$  &   $39.92$  & mr: 12; x: 10\\ 
NGC 3607  &   $8.14$  &   $36.79$  &   $38.79$  & mr: 12; x: 10\\
NGC 3608  &   $8.67$  &   $35.90$  &   $38.20$  & mr: 12; x: 10\\
NGC 3610  &   $8.09$  &   $<35.29$&  $39.05$  & mr: 12; x: 16\\
NGC 3627  &   $7.24$  &   $36.37$  &   $38.30$  & m: 20; r: 11; x: 21\\ 
NGC 3628  &   $7.24$  &   $36.40$  &   $38.51$  & mx: 14; r: 17$^c$\\
NGC 3675  &   $7.1$   &   $<36.3$  &   $<38.1 $  & mrx: 14\\
NGC 3941  &   $7.37$  &   $35.61$  &   $39.27$  & mx: 14; r: 17\\
NGC 4143  &   $8.16$  &   $37.11$  &   $39.97$  & mr: 12; x: 10\\
NGC 4168  &   $8.13$  &   $37.44$  &   $<38.91$  & mrx: 12\\
NGC 4203  &   $7.79$  &   $37.10$  &   $40.09$  & mrx: 17, 14\\
NGC 4216  &   $8.09$  &   $36.58$  &   $38.91$  & mx: 14; r: 22$^c$\\
NGC 4233  &   $8.19$  &   $37.36$  &   $40.26$  & mrx: 12\\
NGC 4261  &   $8.72$  &   $38.62$  &   $41.12$  & mx: 12; r: 23\\
NGC 4278  &   $8.61$  &   $38.35$  &   $40.04$  & mx: 14; r: 11\\
NGC 4365  &   $9.01$  &   $<35.42$  &   $38.32$  & mr: 12; x: 10\\
NGC 4459  &   $7.82$  &   $36.04$  &   $38.82$  & mx: 14; r: 24\\
NGC 4472  &   $9.40$  &   $36.60$  &   $<39.36$  & mrx: 12\\
NGC 4473  &   $7.95$  &   $<35.30$  &   $<38.10$  & mr: 12; x: 10\\
NGC 4477   &  $7.89$   &  $35.90$   &   $39.10$   & m: 1; r: 15$^b$; x: 25\\
NGC 4494  &   $7.68$  &   $36.40$  &   $38.74$  & mx: 12; r: 15$^b$\\
NGC 4501  &   $7.79$  &   $36.28$  &   $38.89$  & mx: 14; r: 26\\
NGC 4552  &   $8.92$  &   $38.23$  &   $39.12$  & mx: 12; r: 10\\
NGC 4564  &   $7.94$  &   $<35.13$  &   $38.52$  & mx: 12; r: 10\\
NGC 4565  &   $7.41$  &   $36.55$  &   $39.85$  & mx: 14; r: 17$^c$\\
NGC 4570  &   $8.03$  &   $<35.23$  &   $38.13$  & mx: 12; r: 10\\
NGC 4594  &   $8.46$  &   $37.85$  &   $40.2$  & mrx: 27, 14\\
NGC 4621  &   $8.40 $  &   $35.1$  &   $37.8$  & mrx: 28\\
NGC 4636  &   $8.33$  &   $36.4$  &   $<38.38$  & mr: 12; x: 20\\
NGC 4649  &   $9.07$  &   $37.48$  &   $38.10$  & m: 14; rx: 27\\
NGC 4697  &   $8.31$  &   $35.00$  &   $37.30$  & mrx: 28\\
NGC 4698  &   $7.57$  &   $35.59$  &   $38.69$  & mrx: 14$^c$\\
NGC 4736  &   $7.05$  &   $35.51$  &   $38.48$  & mx: 14; r: 29\\
NGC 4754  &   $7.76$  &   $<35.31$  &   $38.27$  & mrx: 11\\
NGC 4762  &   $7.63$  &   $36.58$  &   $38.26$  & mrx: 11\\
NGC 4772  &   $7.57$  &   $36.48$  &   $39.30$  & mrx: 14, 17\\
NGC 5576  &   $8.44$  &   $<35.50$  &   $38.88$  & mrx: 12; x: 30\\
NGC 5638  &   $7.60$  &   $<35.52$  &   $<38.33$  & mrx: 12; x: 30\\
NGC 5813  &   $8.75$  &   $37.49$  &   $38.79$  & mr: 12; x: 10\\
NGC 5838  &   $9.06$  &   $36.50$  &   $38.97$  & mrx: 12; x: 10\\
NGC 5845  &   $8.69$  &   $<35.46$  &   $39.05$  & mrx: 12\\
NGC 5846  &   $8.43$  &   $36.62$  &   $39.54$  & mrx: 14\\
NGC 5866  &   $7.81$  &   $37.04$  &   $38.57$  & mx: 14; r: 22\\
NGC 6109  &   $8.56$  &   $39.44$  &   $40.35$  & m: 1; rx: 2\\
NGC 6500  &   $8.28$  &   $39.35$  &   $40.56$  & mrx: 20; x: 30\\ 
NGC 7626   &   $8.71$  &   $38.48$  &   $40.97$  & mrx: 14,  r: 17 \\
IC 1459      &   $ 9.$    &    $39.69$  &  $40.64$  & mrx: 20\\ 
IC 4296     &    $9.1$   &  $38.72$   &   $40.24$  & mrx: 20 
\enddata
\tablerefs{
(1) \citet{Wu2011}; DW15; (2) \citet{Hardcastle2009}; (3) \citet{Gentile2007};
(4) \citet{Bender2005}; (5) \citet{Garcia2010}; (6): \citet{Bosch2010};
(7) \citet{YL2015}; (8) \citet{Devereux2003}; (9) \citet{Miller2010, King2016};
(10) \citet{Pellegrini2010}; (11) \citet{Nagar2001}; (12) \citet{Nyland2016};
(13) \citet{Walsh2013}; (14) \citet{Ho2009}; \citet{HGFS2009};
(15) \citet{Brown2011}; (16) \citet{Miller2012a};  (17) \citet{Nagar2005};
(18) \citet{Wrobel2012}; (19) \citet{Wong2014};
(20) M03; (21) \citet{Grier2011}; (22) \citet{Filho2004}; (23) \citet{LM1997};
(24) \citet{Ho2002}; (25) \citet{Akylas2009} (26) \citet{HU2001};
(27) G09; (28) \citet{Wrobel2008}; (29) \citet{Nagar2002}; (30) \citet{Terashima2003}; (31) \citet{Miller2012b}.
}\label{tab1}
\tablenotetext{a}{the references for black hole mass (labelled `m'), $\lr$ (labelled `r') and $\lx$ (labelled `x').}
\tablenotetext{b}{VLA observed at 1.4 GHz; The flux is convert to that at 5 GHz with a flat radio spectrum assumption, i.e. $\alpha=-0.4$ where $F_\nu \propto \nu^{-\alpha}$.}
\tablenotetext{c}{Only the compact component of Very Long Baseline Array (VLBA) observation (at 8.4 GHz) is considered here. The flux is convert to that at 5 GHz with the assumption $\alpha=-0.4$.}

\end{deluxetable}

\subsection{(Quasi-)$\mbh$-free subsample: the radio/X-ray correlation index $\xi_{\rm X}$}
\label{sec:mbhfreesample}

The key point of the YC05 prediction is the change of the radio/X-ray correlation index $\xi_{\rm X}$. Therefore, the best way to examine this prediction is to study the correlation only between $\lr$ and $\lx$, without the possible ``contamination'' of black hole mass $\mbh$. This is because we have more freedom in the fitting among three quantities $\lr$, $\lx$ and $\mbh$, thus it is difficult to determine the value of $\xi_{\rm X}$ precisely.

Following \citet{deGasperin2011} and DW15 we create a subsample of sources with similar $\mbh$, but has a large dynamical range in X-ray luminosity (in Eddington unit). For simplicity we name it a (quasi-)$\mbh$-free subsample. As shown in Fig.\ \ref{fig:sample}, we select sources with $\mbh$ in the range $10^{8.1} - 10^{9.1}\ \msun$ (one dex in $\mbh$, the boundaries are shown as two red solid curves in this plot). The boundaries are chosen so that the subsample will have largest dynamical range in $\lx/\ledd$, i.e. $10^{\sim -10} < \lx/\ledd < 10^{\sim -5.7}$. This $\mbh$-free subsample includes 42 sources, among which 34 have firm detections in both radio and X-rays. The average black hole mass of this subsample is $\langle\mbh\rangle = 10^{8.71}\ \msun$.

\begin{figure}[htb]
\vspace{0.3cm}
\centering{\includegraphics[scale=0.45]{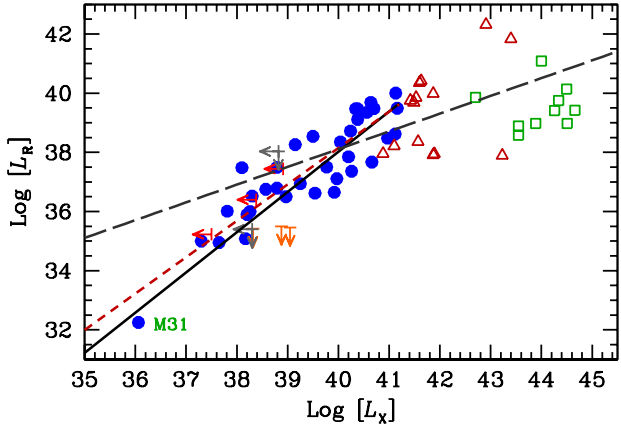}}
\caption{The radio/X-ray correlation of the $\mbh$-free subsample. The black solid curve, with $\xi_{\rm X}=1.36$, represents the fitting of this subsample, cf. Eq.\ \ref{eq:mbhfree}. For comparison, we also show the $\xi_{\rm X}=1.23$ FP of the whole sample (the red dashed curve), and the original M03 FP (the grey long dashed curve), both of which have $\mbh$ fixed to $\langle\mbh\rangle$. The meaning of the symbols is the same as that  in Fig.\ \ref{fig:fp}. In addition, the red open triangles and the green open squares are respectively, the 14 normal LLAGNs selected from DW15 and the 10 AGNs selected from M03, both under the same $\mbh$ constraint. They are shown here as representative of normal LLAGNs that follow the shallower M03 FP.}\label{fig:mbhfree}
\vspace{0.3cm}
\end{figure}

This $\mbh$-free subsample can be used to explore the radio/X-ray correlation, and provide direct constrains on the correlation slope $\xi_{\rm X}$. A linear fit between $\log \lr$ and $\log \lx$ of this $\mbh$-free subsample under the Beyesian method reads (cf. Sec.\ \ref{sec:method}, but note that the quantity $\mbh$ is omitted during the modelling.),
\begin{equation}
\log \lr = 1.36^{+0.07}_{-0.06}\ \log \lx -16.38^{+2.48}_{-2.65}, \label{eq:mbhfree}
\end{equation}
and under the least $\chi^2$ method reads,
\begin{equation}
\log \lr = 1.45^{+0.07}_{-0.07}\ \log \lx -19.59^{+2.83}_{-2.83}.
\end{equation}
We show the Bayesian fitting result as the black solid curve in Fig.\ \ref{fig:mbhfree}. Note that the radio/X-ray correlation slope here is clearly different, at $>7\sigma$ confidence level, from that of the original M03 FP (black long-dashed curve in Fig.\ \ref{fig:mbhfree}, with $\mbh$ set to $\langle\mbh\rangle$). For comparison, we also plot in Fig.\ \ref{fig:mbhfree} the $\xi_{\rm X} = 1.23$ FP of the whole quiescent AGN sample (the red dashed curve), with $\mbh$ also fixed to $\langle\mbh\rangle$. Apparently its difference to Eq.\ \ref{eq:mbhfree} is insignificant.

In Fig.\ \ref{fig:mbhfree}, we additionally show 14 normal LLAGNs (red open triangles) selected from DW15 and the 10 AGNs (green open squares, hereafter the $\mbh$-selected M03 AGNs.) selected from M03, both under the same $\mbh$ constraint. They are shown here as representative of normal LLAGNs and normal AGNs that follow the shallower M03 FP. It is evident from this plot that quiescent AGNs follow a different, steeper, radio/X-ray correlation, compared to the normal LLAGNs and normal AGNs.

Because the physics of accretion and radiation is more relevant to the quantities in Eddington unit rather than its absolute values (such as mass accretion rate and luminosity), we further try to find a linear relationship between $\log (\lr/\ledd)$ and $\log (\lx/\ledd)$ for this $\mbh$-free subsample. Assuming the uncertainty of each quantity is still 0.3 dex, the fitting can be read as,
\begin{equation}
\log (\lr/\ledd)=1.35^{+0.07}_{-0.07}\ \log (\lx/\ledd)+0.49^{+0.53}_{-0.50} \label{eq:eddmbhfree}
\end{equation}
under the Bayesian statistics, and
\begin{equation}
\log (\lr/\ledd)=1.43(\pm0.08)\ \log (\lx/\ledd)+1.13(\pm0.57)
\end{equation}
under the least $\chi^2$ approach. Finally a linear fitting between $\log (\lr/\ledd)$ and $\log (\lx/\ledd)$ of the whole sample can be derived as, $\log (\lr/\ledd)=1.48^{+0.07}_{-0.07}\ \log (\lx/\ledd)+1.13^{+0.51}_{-0.51}$, where the Bayesian approach is adopted. This result is consistent with that derived based on the $\mbh$-free subsample.

\section{Discussions}\label{sec:discussion}

\subsection{Uncertainties of the observational data}\label{sec:questions}


We first discuss the uncertainties of the observational data, i.e. $\sigma_{\rm R}$ and $\sigma_{\rm X}$.

In quiescent AGNs (and normal AGNs also), emission at different wavebands likely correlates with considerable time delays. For one example, \citet{Bell2011} found a $\sim$40-d time delay of radio (at $\sim$5 GHz) to X-rays in the LLAGN NGC~7213. Thus, in order to explore the ``intrinsic'', physically-connected FP, a correction of the radio/X-ray timelag should be applied. However, in practice such correction is almost impossible for AGNs, as it requires long-term intense coordinated monitoring in radio and X-rays on individual sources.\footnote{Note that, the radio/X-ray timelag is considerably small in BHBs. Consequently, it is crucial to use (quasi-)simultaneous radio and X-ray observations to explore the radio/X-ray correlation in BHBs. Additional timelag correction is usually not necessary for the BHB cases.} Moreover, the monitoring should also be able to capture at least one ``outburst'' in each AGN, in order to have a reliable measurement of the timelag. With these obstacles/challenges, to date very few AGNs have such intense monitoring (see e.g. \citealt{Bell2011, King2013} for such monitoring in AGNs). The time interval between radio and X-rays for the data shown in Table\ \ref{tab1} is typically of order of months to years.

The AGNs are variable, with variability amplitude possibly as high as $\sim 100\%$ on time-scales shorter than the time interval of observations \citep{Ho2001}. Such uncertainty dominates over the actual observational uncertainties in the fluxes (or luminosities; $\lr$ and $\lx$) of individual observation reported in literature. Admittedly different estimations will result in somewhat different fitting results \citep{Kording2006}. There are several estimations on the systematical uncertainties of the observed fluxes (e.g. \citealt{Kording2006}, DW15), and we adopt isotropic ones, following M03 and G09. DW15 adopted a slightly smaller scatter in radio luminosities, i.e. $\sigma_{\rm R}=0.2$ dex \citep{Ho2001}, compared to that in X-rays. With this modification to the whole quiescent AGN sample, we carry out Bayesian regression analysis and find that,
\begin{equation}
\xi_{\rm X} = 1.09^{+0.07}_{-0.07}, \hspace{0.5cm} \xi_{\rm M} = 0.70^{+0.12}_{-0.11}, \hspace{0.5cm} c = -11.61^{+2.19}_{-2.25}.
\end{equation}
Note that this result is consistent with a steeper radio/X-ray correlation for quiescent AGNs (cf. Eq.\ \ref{eq:fpfaint}), and disagrees with the claim of a universal FP from sub-Eddington AGNs to quiescent AGNs.

\subsection{impact of Fanaroff-Riley Is and the dimmest M31 on the FP derived}
\label{sec:fr1impact}

It has been known for years that radio-loud sources systematically follow a steeper radio/X-ray correlation (e.g., \citealt{Li2008, deGasperin2011}; DW15). There are 11 sources in our quiescent AGN sample that belong to Fanaroff-Riley (FR, \citealt{FR1974}) Type Is, which are generally radio-loud. One concern is that, the FP derived above (cf. Eq.\ \ref{eq:fpfaint}) may be biased by these sources. To examine the impact of FR Is, we carry out a Bayesian regression analysis to a sub-sample (consist 64 sources) that excludes those FR Is, and the fitting coefficients are,
\begin{equation}
\xi_{\rm X} = 1.18^{+0.08}_{-0.07}, \hspace{0.5cm} \xi_{\rm M} = 0.57^{+0.12}_{-0.11}, \hspace{0.5cm} c = -14.06^{+2.48}_{-2.59}.
\end{equation}
Obviously, it is consistent with the YC05 FP, and disagrees with the M03 FP.

Another concern is that, the FP derived in this work is biased by the dimmest LLAGN in our sample M31 ($\lx/\ledd \sim 10^{-10}$), since the deviation to M03 FP is insignificant for most of the sources in our sample, cf. Fig.\ \ref{fig:fp}. We argue this is actually a misunderstanding, since the radio/X-ray correlation slope can already be determined statistically by abundant other quiescent AGNs. We demonstrate this by applying the Bayesian analysis to a subsample that further excludes M31 (63 sources in total). The correlation coefficients are,
\begin{equation}
\xi_{\rm X} = 1.12^{+0.08}_{-0.08}, \hspace{0.5cm} \xi_{\rm M} = 0.61^{+0.12}_{-0.12}, \hspace{0.5cm} c = -12.01^{+2.69}_{-2.75},
\end{equation}
consistent with the expectation of a YC05 FP in these systems.

\subsection{Comparison with previous works}
\label{sec:otherworks}

Most of the work in literature on the fundamental plane of black hole activity bias towards moderately brighter systems (e.g. M03; \citealt{Falcke2004, Kording2006, Li2008}; G09; \citealt{Fan2016, Nisbet2016}), thus are irrelevant to this work. In this section we discuss the relation of our work to some of the related works published in recent years, i.e. those include the quiescent accretion systems.

\subsubsection{BHBs in their quiescent states: \citet{Gallo2006, Gallo2014, Corbel2008}}\label{sec:bhb}

In our work, we exclude from the sample data points of BHBs in their quiescent states.

Currently there are three BHBs with reported observations in both radio and X-rays during their quiescent states, e.g. A 0620-00 \citep{Gallo2006}, V404 Cyg \citep{Corbel2008} and XTE J1118+480 \citep{Gallo2014}. It is claimed that these sources in their quiescent state follow the extension of the original $\xi_{\rm X}\approx0.6$  FP \citep{Gallo2006, Corbel2008, Gallo2014}. We argue that the claim of original FP down to quiescent states in BHBs is not robust as claimed. The reasons are as follows (see also \citealt{Yuan2014, XY2016}). First, A 0620-00 only have observations in quiescent state, but lacks data in hard state. Secondly, V404 Cyg is still too bright, with $\lx\sim 10^{\sim -6.8}\ledd$, to show clear deviation to the original FP\footnote{Note that, detailed modelling on the quasi-simultaneous multiband (radio up to X-rays) spectrum of V404 Cyg at such low X-ray luminosity indeed supports that the X-ray emission at the quiescent state is of jet origin \citep{Xie2014}.}. Finally, the radio detection of XTE J1118+480 at $\lx \sim 10^{-8.5}\ \ledd$ is admittedly marginal, at 3$\sigma$ level.

Moreover, we note that since the correlation is established in a statistical sense. Few individual sources that do not follow the new correlation can not be taken to argue against the existence and correctness of the new relationship. For example, we note that there also exist some ``outliers'' of the original correlation, as we introduce in {\it Introduction}.

\subsubsection{Sgr A* in quiescent and flare states}
\label{sec:sgra}

\begin{figure*}[htb]
\centering{\includegraphics[scale=0.45]{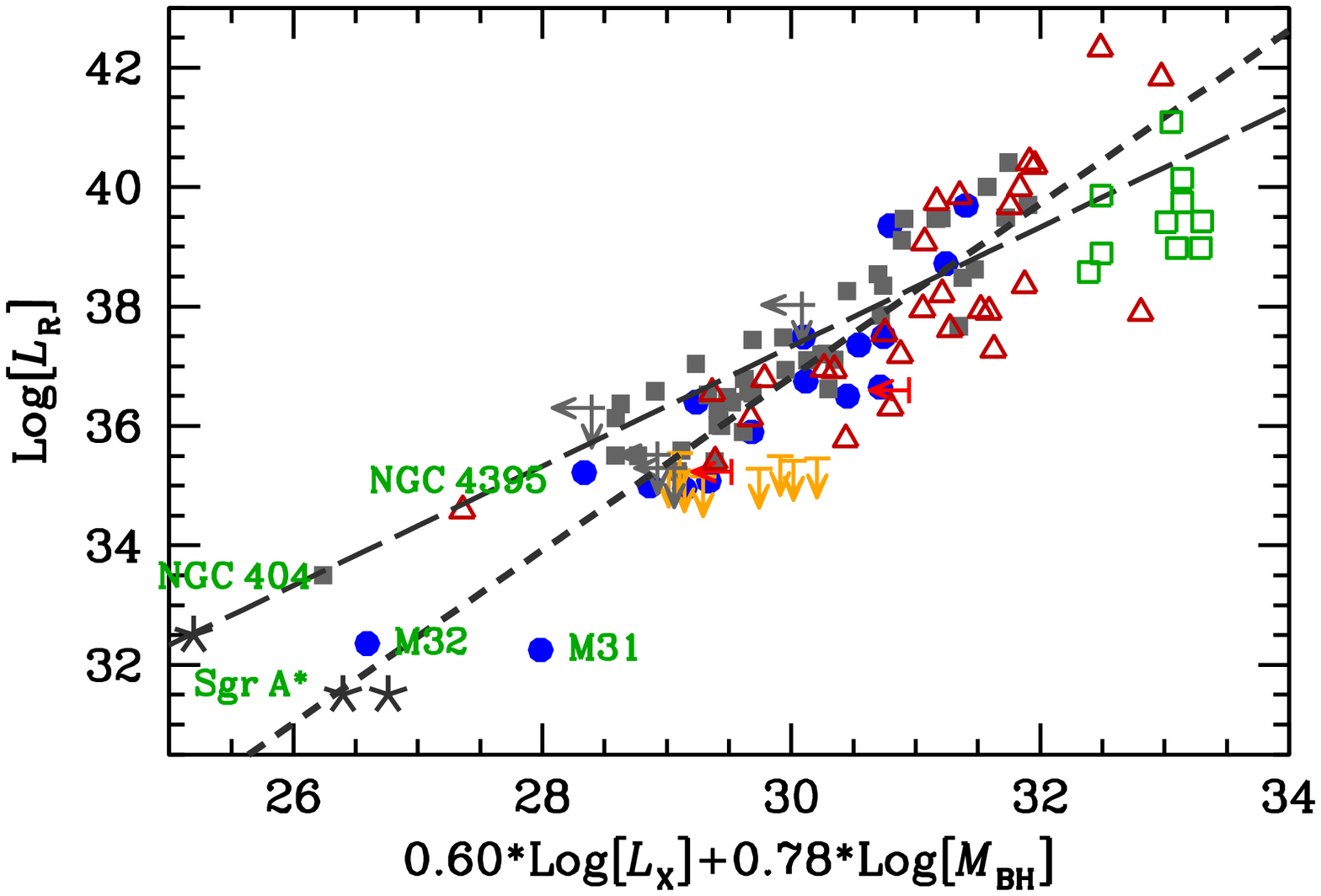}
\hspace{0.4cm}\includegraphics[scale=0.45]{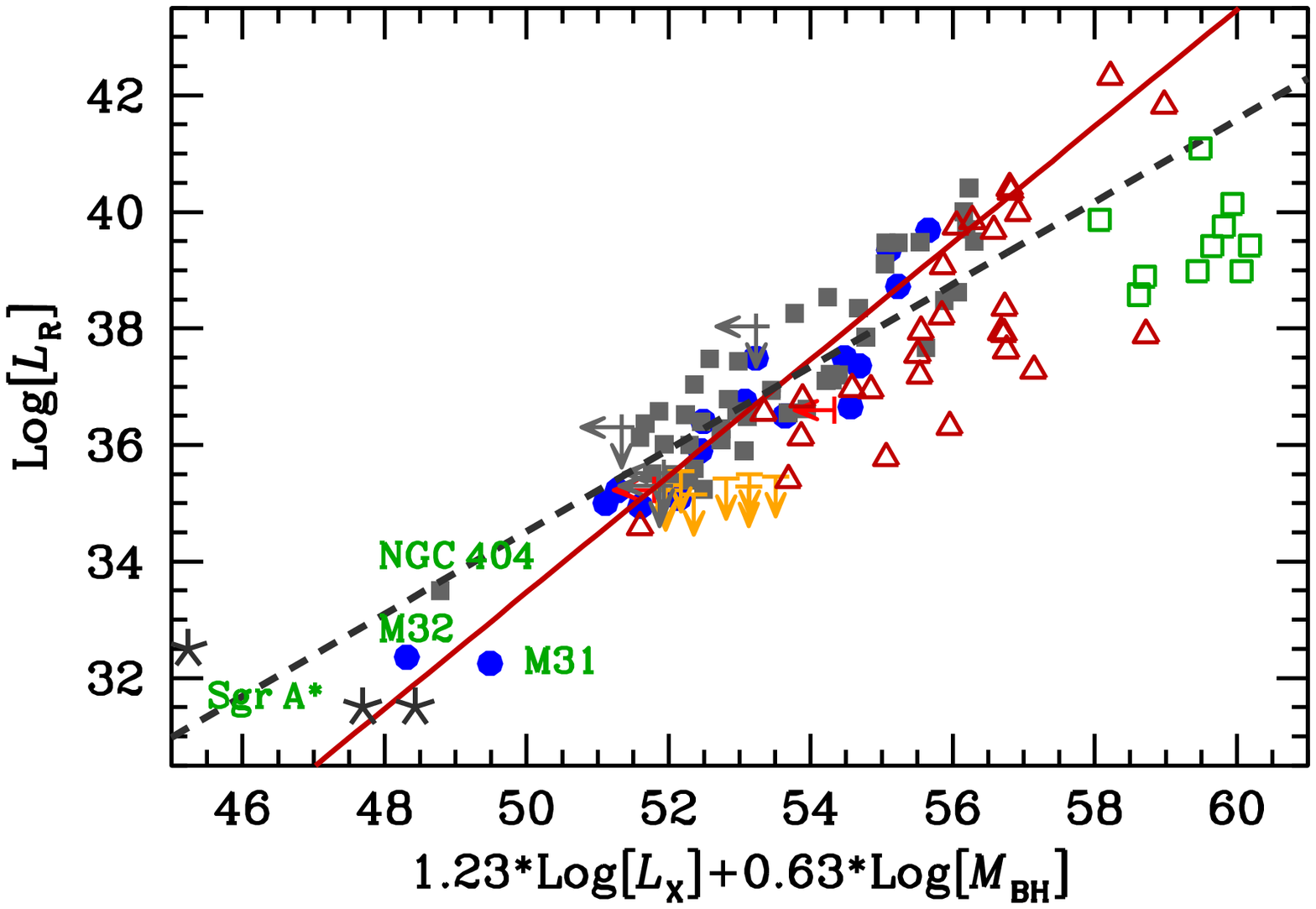}}
\caption{Observational data of the combined sample plot against the M03 FP ({\it left panel}, the grey long dashed curve) and the $\xi_{\rm X}=1.23$ YC05 FP derived in this work ({\it right panel}, the red solid curve). Here the whole sample includes both the quiescent AGNs of this work and also the normal LLAGNs (28 sources from DW15 as representatives; red open triangles). In both panels, symbols and curves are of the same meaning to those in Fig.\ \ref{fig:fp}, with one modification, i.e. the quiescent sources shared by both XY and DW15 are now shown as grey filled squares. In addition, the grey dashed curve shows a joint fitting to this combined sample in both panels, where the black hole mass is fixed to $10^{8.5}\ \msun$. In both panels, the green open squares are the 10 $\mbh$-selected M03 AGNs, cf. Fig.\ \ref{fig:mbhfree}, and they are taken as representative of brighter AGNs.}\label{fig:fp_dw}
\end{figure*}

For the AGN sample in literature, there is one source, i.e. Sgr A*, that deserves special discussion. This source, with $\mbh=4.30\times10^6\ \msun$ \citep{Genzel2010}, has been included in most previous studies on FP. However, as emphasised in Section\ \ref{sec:sample}, there are compelling, independent pieces of evidence against the existence of jet in Sgr A* during its quiescent state (e.g. \citealt{Shen2005, Doe2008, Li2014}), i.e. it does not satisfy the ``existence of jet'' prerequisite in the study of FP thus should be excluded from the sample. During the quiescent state of Sgr A*, the low-frequency radio emission likely originate from the non-thermal electrons of the hot accretion flow, while the X-ray emission is the bremsstrahlung radiation by diffuse gas around $\sim 10^5\ R_{\rm s}$ (see \citealt{Baganoff2003, Yuan2003, Wang2013} and references therein). As shown by the leftmost black asterisk in Fig.\ \ref{fig:fp_dw}, we find that, Sgr A* in its quiescent state agrees with the M03 FP (only a coincidence from our point of view), but disagrees with the new $\xi_{\rm X}\approx 1.23$ FP, as it is more than three orders of magnitude brighter in $\lr$ at the given $\lx$ (see also \citealt{Markoff2005}).

Sgr A* undergoes numerous flares, which are observed in sub-millimeter, infrared and X-rays (e.g., \citealt{Baganoff2003, Neilsen2013} and references therein). These flares usually last 0.1 -- 1 hr \citep{YZ2006, Neilsen2013}, indicating that they are from nuclear regions of Sgr A*. During these flares, the luminosities are enhanced by a factor as much as  $\sim10\%$ in radio (e.g. at $\sim$20 GHz, e.g., \citealt{YZ2006, B2015}) and $\sim$100 ($\sim$400 in extreme cases) in X-rays (2-10 keV, e.g., \citealt{Baganoff2003, Neilsen2013}), compared to the quiescent ``non-flare'' state. We note that, interestingly, these radio and X-ray flares are likely produced from jet \citep{Li2016b, YZ2006, B2015}. If this is indeed the case, then Sgr A* in its flare state should be included in our quiescent AGN sample, and they are qualified to test the FP of quiescent AGNs. This is examined in Figs.\ \ref{fig:fp} and \ref{fig:fp_dw}, where the flare state of Sgr A* is shown by the two black asterisks with $\lr\approx 10^{31.5}\ \ergs$ ($10\%$ that of the quiescent state) and $\lx\approx 10^{35.38}$ and $\approx10^{35.98}\ \ergs$ ($100$ and $400$ times that of the quiescent state), respectively. Here we only take the enhanced luminosities (compared to the ``non-flare'' quiescent state) into account, i.e. only these components are of jet origin. Consistent with our expectation, we find that Sgr A* in its flare state agrees nicely with the steep YC05 FP (cf. Eq.\ \ref{eq:fpfaint} and Fig.\ \ref{fig:fp}), but disagrees with the original M03 FP (cf. Fig.\ \ref{fig:fp_dw}).

\subsubsection{\citet{Yuan2009} (YYH09)}

YYH09 is the first to discover from observational data the existence of a new FP, with $\xi_{\rm X}\approx 1.22$, in well agreement with the theoretical prediction by YC05. During their sample compilation, YYH09 exclude data from normal LLAGNs, and only include data from quiescent AGNs. Besides, Sgr A* is also excluded from their sample, for reasons listed above. Our work is a natural extension of YYH09. The main improvement to YYH09 is the sample size, which is enlarged from 22 to 75. Besides, the radio emission of several sources, e.g., M31 (\citealt{Bender2005, Garcia2010}, $\lx\sim 10^{-10.2}\ \ledd$) and M32 \citep{YL2015}, is now firmly detected, while in YYH09 their radio emission only has upper limit constraint. The main conclusion remains unchanged, i.e. our enlarged sample confirms the discovery of YYH09.

\subsubsection{\citet{Dong2015} (DW15)}

DW15 recently select from the (nearly-)complete flux-limited Polmar survey a sample of 72 AGNs, and combine with a large sample of observational data points of BHBs, to investigate the FP jointedly from sub-Eddington to quiescent systems. They claim those with $\lx\lesssim L_{\rm X,crit}$ (24 AGNs under a detailed definition of $L_{\rm X,crit}$) seem to follow the original M03 FP, which disagrees with this work (and YYH09 also).

There are mainly two reasons on such discrepancy. First, notable differences between their sample and ours are observed. Their sample includes the BHBs in their hard and quiescent states, and also Sgr A* in quiescent state. Moreover, because their AGN sample is limited to data from the Polmar survey, the quiescent AGN subsample of DW15 lack sufficient number of sources whose X-ray Eddington ratios are sufficiently low. For the $\lx/\ledd<10^{-8}$ regime, there are only three sources in their sample (except Sgr A*), while there are 17 sources in our sample. Several notable faint sources are missed, e.g., M31 (\citealt{Bender2005, Garcia2010}, $\lx\sim 10^{-10.2}\ \ledd$) and NGC~3115 (\citealt{Wrobel2012, Wong2014}, $\lx < 10^{-9.5}\ \ledd$). These faintest sources, with possibly largest deviation to the original FP, are of crucial importance to reveal the new trend. Secondly, they mainly provide a joint fitting of both normal LLAGNs and quiescent AGNs\footnote{We note that, since their quiescent subsample also includes data points from BHBs and Sgr A*, a fitting of such subsample, as shown in the bottom-right panel of Fig. 1 in DW15, will be misleading also. Indeed, as shown in their Fig. 2, the subsample of quiescent AGNs with similar BH mass do hint on a much steeper radio/X-ray relationship.}, while we focus on the quiescent AGNs only here. As illustrated below in Section\ \ref{sec:jointfit}, we argue that the small number of quiescent sources in their sample and a joint fitting are the reasons for the discrepancy between their result and ours.

\subsection{Disadvantages of a joint fitting of both normal LLAGNs and quiescent AGNs}
\label{sec:jointfit}

We here discuss the disadvantages of a joint fitting of a sample that includes both normal LLAGNs and quiescent AGNs, especially when the steeper correlation is statistically insignificant as the sample lack sufficient number of dimmest sources. For this purpose, we consider a combined AGN sample, including not only the quiescent AGNs of this work (hereby named as the XY sample) but also 28 normal LLAGNs of DW15 (red open triangles in Fig.\ \ref{fig:fp_dw}. Note that, except for Sgr A*, the rest 44 AGNs are selected into the XY sample). We note that it has been known for years that some sources do not follow the original M03 FP (e.g., \citealt{Li2008, deGasperin2011} for AGNs, and \citealt{Coriat2011} for BHBs. cf. {\it Introduction}). Consequently, the exact values of the fitting depend on both sample selection and statistical methods, as noted in \citet{Kording2006} and demonstrated in Sec.\ \ref{sec:results}.

We analysis the combined sample through the Bayesian approach. The correlation indexes now read as,
\begin{equation}
\xi_{\rm X} = 0.87^{+0.04}_{-0.04}, \hspace{0.5cm} \xi_{\rm M} = 1.29^{+0.09}_{-0.08}, \hspace{0.5cm} c = -8.05^{+1.48}_{-1.50}. \label{eq:fp_w_dw}
\end{equation}
Note that the value of $\xi_{\rm X}$ is between the result of DW15 and the present work shown in Section\ \ref{sec:results}. This indicates that the value of $\xi_{\rm X}$ is somewhat sensitive to the fraction of quiescent sources included in the sample, more fraction of quiescent ones will make its value larger and the radio/X-ray correlation slope steeper. This is consistent with our expectation.

Now we do the statistical analysis by another way. We fix the value of $\xi_{\rm X}$ and test two-parameter fittings of the combined sample. We find that both the $\xi_{\rm X}\approx 0.6$ (M03-like) correlation and the $\xi_{\rm X} \approx 1.23$ (YC05-like) correlation could provide almost equally good (or bad) fits to the combined sample, i.e. both of which leads to relatively similar reduced $\chi^2$ values. Consequently it is difficult to judge from statistical instead of physical point of view which fitting is better.

Fig.\ \ref{fig:fp_dw} shows the data points of the combined sample plotted against the M03 FP (left panel; cf. M03) and the YC05 FP (right panel; cf. Eq.\ \ref{eq:fpfaint}). For clarity, the quiescent sources shared by both XY and DW15 are now shown as grey filled squares. For comparison, we show in the two panels the fitting of this combined sample (Eq.\ \ref{eq:fp_w_dw}) as grey dashed curve, where $\mbh$ is set to $10^{8.5}\ \msun$. Based on the above fitting results as well as the direct comparison between the two panels, we can understand the reason for the discrepancy between this work (and other similar ones) and DW15 (and other similar ones). The first reason has been pointed out already below Eq.\ \ref{eq:fp_w_dw}, i.e., the correlation index will be determined by the fraction of quiescent sources (or equivalently normal sources) whose $\lx$ are far away from $L_{\rm X, crit}$ included in the sample. The second reason is that, as shown in Fig.\ \ref{fig:fp_dw}, because the luminosity scales as $L_{\rm R, X}\propto L_{\rm R, X}/\ledd \times \mbh$ and the differences in $\mbh$ is sufficiently large (by $\sim$3 orders of magnitude) in both quiescent and normal low-luminosity AGNs, the quiescent sources will mix up with those normal ones in $L_{\rm X}$ and $L_{\rm R}$. Consequently, quiescent sources that are not dim enough (in $\lx/\ledd$) will only contribute to the scatter of the original M03 FP, and the YC05 FP will be obscured.

\begin{figure}[htb]
\centering{\includegraphics[scale=0.45]{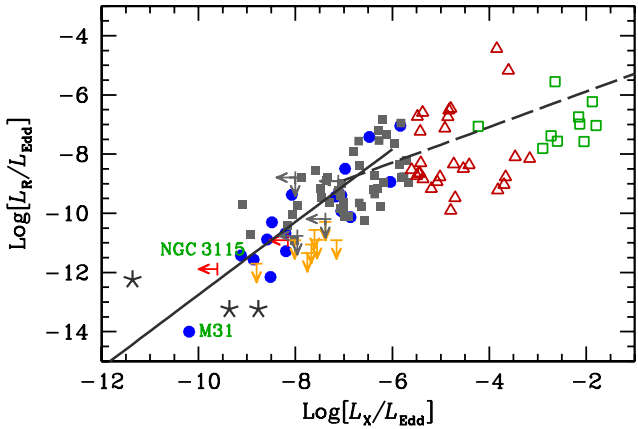}}
\caption{The $\lr/\ledd$ -- $\lx/\ledd$ relationship of the combined sample. The meaning of the symbols is the same as that in Fig.\ \ref{fig:fp_dw}. The solid and the dashed curves represent, respectively, the YC05 FP (cf. Eq.\ \ref{eq:fpfaint}) and the original M03 FP, where the black hole mass is fixed to $10^{8.5}\ \msun$.}\label{fig:fpedd}
\end{figure}

We argue that since the YC05 prediction is based on the difference in $\lx/\ledd$ rather than $\lx$, the correct way to examine this prediction is to investigate the sample only consisting of sources with $\lx/\ledd \la 10^{-6}$. This is also more physical since the underlying physics is determined by $\lx/\ledd$ instead of $\lx$. A $\mbh$-free subsample (cf. Section\ \ref{sec:mbhfreesample}) will help to solve this problem. Indeed, from the $\mbh$-selected sample, the quiescent AGNs do follow a radio/X-ray correlation that is much steeper compared to that followed by those normal (LL)AGNs. This result is clearly shown in Fig.\ \ref{fig:mbhfree}.

Based on this consideration, we propose that a better way to analyze the correlation is to replace $\lr$ and $\lx$ with $\lr/\ledd$ and $\lx/\ledd$, and investigate the FP under a revised three-dimensional ($\log(\lr/\ledd)$, $\log(\lx/\ledd)$, $\log(\mbh)$) space. One notable advantage of this new space is that, objects with different Eddington ratios will be separated automatically. For a demonstration of this advantage, we show in Fig.\ \ref{fig:fpedd} the $\lr/\ledd$ -- $\lx/\ledd$ relationship of the combined sample, with additional 10 $\mbh$-selected M03 AGNs. From this plot, we can see clearly that the correlation slope above and below $\lx/\ledd\sim10^{-6}$ is different. The quiescent sources do follow a steeper radio/X-ray correlation compared to that of normal LLAGNs, in agreement with YC05 and our finding in the present paper. This new parameter space will be very helpful to the investigation of FP at different luminosity regimes (in Eddington unit), where a change in the FP may be observed, as a consequence of the change in accretion mode at that luminosity regime (e.g. \citealt{XY2012, XY2016, Yuan2014, Yang2015}).

As a preliminary test, we consider our whole quiescent AGN sample (the XY sample). The scatter of each quantity is fixed to 0.3 dex, for the sake of simplicity. Under the Bayesian approach we find that,
\begin{eqnarray}
\log (\lr/\ledd)& =& 1.29^{+0.07}_{-0.06}\ \log (\lx/\ledd) \nonumber\\
& & + 0.89^{+0.10}_{-0.09}\ \log \mbh - 7.61^{+0.87}_{-0.89}.
\end{eqnarray}
The scatter in $\lr/\ledd$ is 0.82 dex. This result can be re-written as $\lr\propto \lx^{1.29}\mbh^{0.60}$.

\subsection{The distribution of Radio-loudness parameter $R_{\rm X}$ versus Eddington ratio in faint AGNs with $\lx/\ledd\lesssim 10^{-6}$}
\label{sec:R_lbol}

It is widely known that LLAGNs tend to be radio-loud systematically, and the radio-loudness parameter, defined as $R_{\rm X}=\lr/\lx$, scales inversely with Eddington ratios $L_{\rm bol}/\ledd$ (among others see e.g., \citealt{Ho2008, Nyland2016}), where $L_{\rm bol}$ is the bolometric luminosity. We note that the FP can be re-written as, $R_{\rm X}\propto L_{\rm X}^{\xi_{\rm X}-1}$. If $L_{\rm bol}$ scales positively with $\lx$ (likely a reasonable assumption), then the YC05 FP, with $\xi_{\rm X}\approx$ 1.23$>1$, predicts that quiescent AGNs follow a positive $R_{\rm X}$ -- $L_{\rm bol}/\ledd$ relationship, opposite to those normal LLAGNs.

\begin{figure}[htb]
\centering{\includegraphics[scale=0.45]{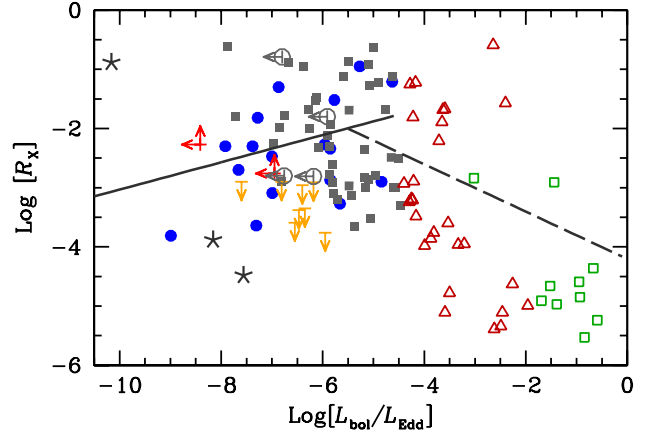}}
\caption{Distribution of radio-loudness parameter $R_{\rm X}$ versus Eddington ratio $L_{\rm bol}/\ledd$, where the bolometric luminosity is estimated as, $L_{\rm bol} = 16\ \lx$. The meaning of the symbols is the same as that in Fig.\ \ref{fig:fp_dw}. Note that those black open circles are sources whose $\lx$ and $\lr$ are upper-limit constraints, i.e. their $R_{\rm X}$ is actually un-constrained. The solid and the dashed curves represent, respectively, the YC05 FP (cf. Eq.\ \ref{eq:fpfaint}) and the original M03 FP, where the black hole mass is fixed to $10^{8.5}\ \msun$. }\label{fig:R_lbol}
\end{figure}

Fig.\ \ref{fig:R_lbol} illustrates the distribution of radio-loudness parameter $R_{\rm X}$ versus Eddington ratios for both quiescent AGNs and normal LLAGNs. For normal LLAGNs, we take those from DW15 as representatives. Following \citet{Ho2008}, we estimate the bolometric luminosity simply from the X-ray luminosity, i.e. $L_{\rm bol}=16\ \lx$. Note that those black open circles are sources whose $\lx$ and $\lr$ are upper-limit constraints, i.e. their $R_{\rm X}$ is actually un-constrained. The black solid and long-dashed curves, respectively, show the results derived from the YC05 FP and the original M03 FP, where the BH mass is fixed to $10^{8.5}\ \msun$. From this figure, the possible turnover at Eddington ratio $L_{\rm bol}/\ledd = 10^{\sim -5.5}$ (or equivalently $\lx/\ledd = 10^{\sim -6.5}$; see also \citealt{Yang2015} for this value, as constrained by the X-ray spectral properties in AGNs and BHBs.) is admittedly less evident. More observations of quiescent AGNs with $\lx/\ledd \lesssim 10^{-8}$ (or Eddington ratio $L_{\rm bol}/\ledd \lesssim 10^{-7}$) are urged to examine this new trend in future.

\section{Summary}
\label{sec:summary}
The Fundamental Plane provides a direct evidence on the disk-jet connection (e.g. M03; \citealt{Falcke2004, Merloni2006}; YYH09). One remaining question under active debate is whether or not those very faint accretion systems (i.e., $\lx/\ledd$ below a critical value $\sim 10^{-6}$) follow the original M03 FP or the steeper (in sense of the radio/X-ray correlation slope) YC05 relationship. Many works favour the YC05 FP (e.g., \citealt{Wu2007, Pellegrini2007, Wrobel2008}; YYH09; \citealt{deGasperin2011,  Younes2012, Reynolds2014}), while several others favour an universal FP extending down to quiescent/faint systems (\citealt{Gallo2006, Gallo2014, Corbel2008}; DW15). In this work we re-visit this problem. For this aim, the quality of the data (mainly the radio data) of BHBs is not satisfactory thus the conclusion based on that is not convincing, as we argue in Section\ \ref{sec:bhb}. Therefore we focus on quiescent AGNs, paying special attention to the radio/X-ray correlation slope $\xi_{\rm X}$. Compared to previous studies, in our work we gather from literature as many faint AGNs as possible thus our sample is the largest to date, with 5 sources fainter than $10^{-9} \ledd$ in X-rays (Fig.\ \ref{fig:sample}). As we show in the paper, these faint sources are crucial to discriminate different correlations. Our main results are summarized as follows,

\begin{itemize}

\item  Based on our quiescent AGN sample, we find that quiescent AGNs follow a steeper FP compared to M03 FP. The radio/X-ray correlation slope $\xi_{\rm X}\approx 1.23$, in good consistency with the prediction of YC05 (Fig.\ \ref{fig:fp}).

\item To further focus on the radio/X-ray correlation but eliminate any possible contamination of the black hole mass, we create a sub-sample in which the black hole mass is similar. For such a $\mbh$-free sub-sample, we find that the value of $\xi\approx 1.36$ (Fig.\ \ref{fig:mbhfree}).

\item We have further explored the reasons for the discrepancy between the present result and some previous ones. We find that, for the combined AGN sample that includes sources of both $\lx/\ledd\la10^{-6}$ and $\lx/\ledd\ga 10^{-6}$, the value of $\xi_{\rm X}$ is $\approx 0.87$ (Eq.\ \ref{eq:fp_w_dw}), which is between 0.6 and $\sim$1.2--1.3. It is expected that the exact value of $\xi_{\rm X}$ in general will be determined by the fraction of quiescent (or equivalently normal) sources in the sample. Put it in another way, we find that the $\xi_{\rm X}\approx 0.6$ correlation and the $\xi_{\rm X} \approx 1.23$ correlation provide almost equally good (or bad) fits to the combined sample, and it is difficult to judge from statistical instead of physical point of view which fitting is better. We thus argue this approach is not the best way to examine the correlation at quiescent regime. In the traditional approach, unless the quiescent sources are extremely faint (in $\lx/\ledd$), we cannot separate them from normal ones, since they are mixed up with in $\lx$ and $\lr$ due to the large range in black hole mass. Consequently, quiescent sources will only contribute to the scatter of the original FP (the left panel of Fig.\ \ref{fig:fp_dw}).

\item Given the above reasons, we propose that a better way to investigate the Fundamental Plane is to use a revised three-dimensional space spanned by $\log(\lr/\ledd)$, $\log(\lx/\ledd)$, and $\log(\mbh)$. Physically, parameters $\lr/\ledd$ and $\lx/\ledd$ have more direct connections to the accretion/jet physics. One notable advantage of this new space is that, objects with different Eddington ratios (thus may relate to different accretion regimes) are separated automatically. As shown in the Figs.\ \ref{fig:mbhfree}\&\ref{fig:fpedd}, there is clearly a ``break'' in the $\lr/\ledd$ -- $\lx/\ledd$ correlation at a critical luminosity $\lx/\ledd\sim 10^{-6}$, below which the correlation is steeper, consistent with YC05 and YYH09.

\end{itemize}

\section*{Acknowledgements}
We thank Drs. Qingwen Wu and Ai-Jun Dong for helpful discussions and comments, and the referee for valuable suggestions that improve our analysis. FGX thanks Drs. Zhaoming Gan, Doosoon Yoon and Zhen Yan for the help on {\texttt Python}, and Drs. Zhao-Zhou Li and Shi-Yin Shen for the help on Bayesian regression analysis. FGX and FY are supported in part by the National
Program on Key Research and Development Project of China (Grant Nos. 2016YFA0400804 and 2016YFA0400704), the Youth Innovation Promotion Association of Chinese Academy of Sciences (CAS) (id. 2016243), the Natural
Science Foundation of China (grants 11573051, 11633006 and 11661161012), the Key Research Program of Frontier Sciences of CAS (No. QYZDJ-SSW-SYS008). This work has made extensive use of the NASA/IPAC Extragalactic Database (NED), which is operated by the Jet Propulsion Laboratory, California Institute of Technology, under contract with the National Aeronautics and Space Administration (NASA).

\end{document}